\documentclass[preprint,preprintnumbers,amsmath,amssymb,floatfix]{revtex4}

\renewcommand{\section}[1]{\vspace{.2in} \noindent{\Large{{\bf {#1}}}}}

\usepackage{graphicx}
\usepackage{dcolumn}
\usepackage{bm, amsmath, amssymb}

\newcommand{\omz}{\omega_z}
\newcommand{\omc}{\omega_c}
\newcommand{\omcp}{\omega_c^\prime}
\newcommand{\dca}{\Delta_{ca}}
\newcommand{\dpc}{\Delta_{pc}}

\newcommand{\nbar}{\bar{n}}

\newcommand{\snn}{S_{nn}}
\newcommand{\snnplus}{S_{nn}^{(+)}}
\newcommand{\snnminus}{S_{nn}^{(-)}}
\newcommand{\snnplusminus}{S_{nn}^{(\pm)}}

\newcommand{\zho}{z_{\mbox{\scriptsize{ho}}}}
\newcommand{\Zho}{Z_{\mbox{\scriptsize{ho}}}}
\newcommand{\Pho}{P_{\mbox{\scriptsize{ho}}}}
\newcommand{\neff}{N_{\mbox{\scriptsize{eff}}}}
\newcommand{\zbari}{\bar{z}_i}
\newcommand{\ahat}{\hat{a}}
\newcommand{\adag}[0]{a^{\dagger}}

\newcommand{\rfs}{R_{\rm{fs}}}

\newcommand{\rcav}{R_{\rm{c}}}

\newcommand{\comment}[1]{}
\newcommand{\hami}{\mathcal{H}}

\newcommand{\bdag}{b^{\dagger}}
\newcommand{\omone}{\omega_{1}}
\newcommand{\omtwo}{\omega_{2}}
\newcommand{\nmax}{n_{\mbox{\scriptsize{max}}}}

\newcommand{\cin}{c_{in}}
\newcommand{\cindag}{c_{in}^{\dagger}}
\newcommand{\din}{d_{in}}
\newcommand{\dindag}{d_{in}^{\dagger}}
\newcommand{\cout}{c_{out}}

\newcommand{\dout}{d_{out}}
\newcommand{\doutdag}{d_{out}^{\dagger}}

\begin{document}

\title{Observation of quantum-measurement backaction with an ultracold atomic gas}

\author{Kater W. Murch$^{1}$}
\author{Kevin L. Moore$^{1}$}
\author{Subhadeep Gupta$^{1}$}
\author{Dan M. Stamper-Kurn$^{1,2}$}
\affiliation{
    $^1$Department of Physics, University of California, Berkeley CA 94720 \\
    $^2$Materials Sciences Division, Lawrence Berkeley National Laboratory, Berkeley, CA 94720}
\date{}


\maketitle

\textbf{Current research on micro-mechanical resonators strives for
quantum-limited detection of the motion of macroscopic objects
\cite{schw05phystoday}.  Prerequisite to this goal is the
observation of measurement backaction consistent with quantum
metrology limits \cite{brag95qmbook}. However, thermal noise
presently dominates measurements and precludes ground-state
preparation of the resonator.  Here we establish the collective
motion of an ultracold atomic gas confined tightly within a
Fabry-Perot optical cavity \cite{gupt07nonlinear,colo07,bren07cqed}
as a system for investigating the quantum mechanics of macroscopic
bodies. The cavity-mode structure selects a single collective
vibrational mode that is measured by the cavity's optical
properties, actuated by the cavity optical field, and subject to
backaction by the quantum force fluctuations of this field.
Experimentally, we quantify such fluctuations by measuring the
cavity-light-induced heating of the intracavity atomic ensemble.
These measurements represent the first observation of backaction on
a macroscopic mechanical resonator at the standard quantum limit.}

Various types of micro-mechanical resonators, including singly
\cite{klec06cooling,pogg07} or doubly
\cite{naik06back,giga06cooling,arci06} clamped nanofabricated beams,
thin membranes \cite{thom07mem}, and toroidal structures
\cite{schl06cooling}, have been fabricated and used to study
small-amplitude vibrations. With resonance frequencies in the kHz to
MHz range -- an exception being the GHz resonator of Ref.\
\cite{huan03ghz} -- these resonators remain significantly perturbed
by thermal noise at cryogenic temperatures. Nevertheless, powerful
schemes to cool a single mechanical mode of the resonator below its
ambient temperature have been demonstrated
\cite{arci06,klec06cooling,giga06cooling,naik06back,schl06cooling,thom07mem,pogg07}.
These schemes use either active feedback or the passive dynamical
backaction on a driven resonator, the latter being equivalent to
cavity-induced laser cooling of atoms \cite{hora97,vule00}. The use
of these schemes to achieve ground-state cooling has been discussed
\cite{marq07sideband,wils07groundstate,vita03josabreview}.

In this work, we demonstrate that the collective motion of a trapped
macroscopic ensemble of ultracold atoms may serve as the resonator
for the study of quantum micro-mechanics.  In contrast with the
mechanical systems discussed above, such atoms may be cooled
directly to the ground state of motion. Non-classical states of
motion have been engineered in atomic ensembles \cite{bouc99}, and
the oscillatory motion of an atomic gas has been used to measure
weak forces \cite{harb05casimir}, analogous to measurements using
microfabricated cantilevers \cite{mohi98}. However,  previous
efforts have lacked the means to measure the motion of an atomic
ensemble at the quantum limit.

High-finesse optical cavities have been used to sense the motion of
single atoms \cite{mabu96,hood00micro}. Their sensitivity results
from the spatial variation of the atom-cavity coupling frequency; in
a near-planar Fabry-Perot cavity, this frequency varies as $g(z) =
g_0 \sin k_p z$ along the cavity axis, where $k_p$ is the wavevector
of light near the cavity resonance. In the case where the detuning
$\dca = \omc - \omega_a$ between the bare-cavity (no atoms present)
and the atomic resonance frequencies is large ($|\dca| \gg
\{g_0,\Gamma\}$), a single atom of half-linewidth $\Gamma$ at
position $z$ causes the cavity resonance to be shifted by $g^2(z) /
\dca$.   Measuring the cavity resonance thus provides information on
the atom's position.

Such a measurement may be applied also to monitor the motion of an
ensemble of $N$ atoms that are optically trapped within the
resonator mode. In this case, a single collective degree of freedom
couples exclusively  to a single mode of the cavity (see
supplemental information). For small displacements of the atoms from
their potential minima, we define a collective position operator $Z
= (\neff)^{-1} \sum_i \sin(2 k_p \zbari) \delta z_i$, and the
conjugate momentum $P = \sum_i \sin(2 k_p \zbari) p_i$, with
$\zbari$ being the equilibrium position of the $i^{th}$ atom,
$\delta z_{i}$ its position deviation from equilibrium operator, and
$p_i$ being its momentum.  The cavity then serves to monitor a
specific collective mode of motion in the atomic ensemble, with the
cavity resonance being shifted by $\Delta_N - \neff f_0 Z/ \hbar$
where $\Delta_N = \sum_i g^2(\zbari)/\dca$ is the cavity frequency
shift with all atoms localized at their potential minima and $f_{i}
= -\hbar \partial_z g^2(\zbari)/\dca = f_{0} \sin (2 k_{p}\zbari)$
is the optical dipole force from a single cavity photon.  That is,
the collective mode sensed by the cavity is equivalent to the
center-of-mass motion of $\neff = \sum_i \sin^2(2 k_p \zbari)$ atoms
trapped at locations of maximum sensitivity of the cavity properties
to the atomic position.

With the identification of the collective variables $Z$ and $P$, we
may draw directly on results obtained for the motion of
radiation-pressure-driven mechanical resonators within optical
cavities. For example, we conclude that optical dipole forces in a
driven cavity will displace the collective variable $Z$, shifting
the cavity resonance frequency and leading to cavity optical
nonlinearity and bistability
\cite{dors83bistability,gupt07nonlinear}.   We find also that force
fluctuations arising from the quantum fluctuations of the
intracavity optical field disturb the collective momentum $P$ and
constitute the quantum backaction for cavity-based measurements of
the displacement $Z$ \cite{cave81}.

To assess the impact of these dipole force fluctuations, we consider
the dynamics of the atoms-cavity system with the cavity continuously
driven by laser light of fixed detuning $\dpc$ from the bare-cavity
resonance. The average optical force of $\nbar$ cavity photons
displaces the collective position variable by $\Delta Z = (\hbar k
g_0^2 / m \omz^2 \dca) \nbar$ and thereby shifts the cavity
resonance frequency to $\omcp = \omc + \Delta_N - \neff f_0 \Delta Z
/ \hbar$, where $\omz$ is the trap frequency.   We define collective
quantum operators $a$ and $a^\dagger$ through the relations $Z -
\Delta Z = \Zho (a^\dagger + a)$ and $P = i \Pho (a^\dagger - a)$,
with $\Zho =\sqrt{\hbar/2 m \omz \neff}$ and $\Pho = \hbar / (2
\Zho)$.  As discussed in the supplemental information, we obtain
equations of motion for $a$ and for the cavity field operator $b$ as
\begin{eqnarray}
\frac{d a}{d t} & = & - i \omz a + i \kappa \epsilon (n
- \nbar), \label{eq:dadt}\\
\frac{d b}{d t} & = & -i \omcp b + i \kappa \epsilon(\adag + a) b -
\kappa b + \sqrt{2 \kappa} b_{in}, \label{eq:dbdt}
\end{eqnarray}
where $\kappa$ is the decay rate of the cavity field, $b$ is the
cavity photon annihilation operator and $b_{in}$ represents the
coherent-state input field that drives the cavity.

Here, we introduce a dimensionless ``granularity'' parameter,
$\epsilon = \neff f_0 \Zho / (\hbar \kappa)$, that quantifies the
coupling between quantum fluctuations of the collective atomic and
optical fields. In the non-granular regime, defined by $\epsilon \ll
1$, the generally complex atoms-cavity dynamics described by Eqs.\
\ref{eq:dadt} and \ref{eq:dbdt} are vastly simplified.  To
characterize this regime, consider the impulse $\neff f_0 / 2
\kappa$ imparted upon the collective motion by the single photon
optical force over the $(2 \kappa)^{-1}$ lifetime of a cavity
photon.  For $\epsilon \ll 1$, this impulse is smaller than the
zero-point momentum fluctuations of rms magnitude $\Pho$; thus, the
effects of optical force fluctuations on the atomic ensemble are
adequately described by coarse graining. Likewise, the transient
displacements induced by this impulse will shift the cavity
resonance by an amount that is much smaller than $\kappa$; thus, the
quantum fluctuations of the cavity-optical field are the same as in
the absence of the intracavity atomic gas, with the spectral density
of photon number fluctuations being $\snn(\omega) = 2 \nbar \kappa
(\kappa^2 + (\Delta + \omega)^2)^{-1}$ \cite{marq07sideband} with
$\Delta = \dpc - (\omcp - \omc)$ being the probe detuning from the
atoms-shifted cavity resonance.

We then find the occupation number of the collective atomic
excitation to vary as
\begin{equation}
\frac{d}{d t} \langle a^\dagger a \rangle = \kappa^{2} \epsilon^2
\left[\snnminus + \left(\snnminus - \snnplus\right) \langle
a^\dagger a \rangle \right] \label{eq:adaga}
\end{equation}
where $\snnplusminus = \snn(\pm \omz)$ and we assume $\langle
a^\dagger a \rangle$ remains small. The collective atomic motion is
subject to momentum diffusion, which heats the atomic gas at a
per-atom rate of $\rcav = \hbar \omz \kappa^{2} \epsilon^2 \snnminus
/ N$, and also to coherent damping or amplification of the atomic
motion \cite{hora97,vule00,marq07sideband,wils07groundstate}.

So far we have neglected the force fluctuations on the atoms
associated with incoherent scattering.  As in free space,
spontaneous emission by atoms driven by laser light leads to
momentum diffusion due to both recoil kicks and fluctuations of the
optical dipole force \cite{dali85,gord80}.  Allowing the trapped
atoms to be distributed evenly along the cavity axis, the total
light-induced per-atom heating rate becomes $R = \rfs + \rcav$ where
$\rfs = (f_0^2 / 2 m) (\nbar / \kappa) (1/C)$ is the free-space
diffusive heating rate in a standing wave of light
\cite{gord80,dali85}.  For this atomic distribution, $\neff = N/2$
and we obtain $\rcav = \rfs \times C (1 + (\Delta -
\omz)^2/\kappa^2)^{-1}$.  Thus, in the strong coupling regime of
cavity quantum electrodynamics, with single-atom cooperativity $C =
g_0^2 / 2 \kappa \Gamma \gg 1$, diffusive heating may be dominated
by backaction heating ($\rcav$) for probe frequencies near the
cavity resonance ($|\Delta - \omega_{z}| < \kappa$).

\begin{figure}[tb]
\includegraphics[angle = 0, width = .5\textwidth]{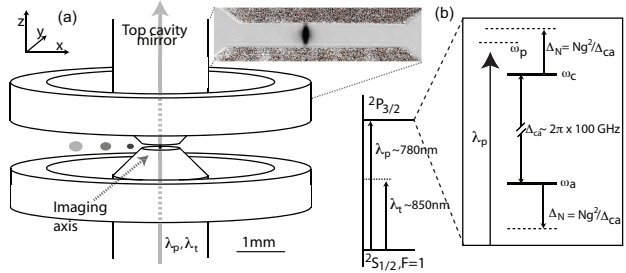}
\caption{(a) Ultracold atoms are produced in a magnetic trap, formed
using electromagnets coaxial with the vertically oriented
high-finesse cavity, and delivered to the cavity center.
Trapping/locking light ($\lambda_t\! =\! 850$ nm) and probe light
($\lambda_p\! =\! 780$ nm) are sent through the cavity and monitored
in transmission.  An absorption image, obtained using probe light
along the $\hat{y}$ axis, shows atoms trapped optically within the
cavity volume. (b) Energy level scheme for the far-detuned
($\Delta_{ca} \gg \sqrt{N} g_{0}$) cavity.} \label{fig:scheme}
\end{figure}

In our experiment, this backaction heating was measured
bolometrically. Because the mechanical $Q$ of the collective
vibrational mode is low ($\sim 40$ as determined in Ref.\
\cite{gupt07nonlinear}), backaction induced excitation of this mode
soon leads to a rise in the total thermal energy of the atomic
sample.  We quantify this energy increase by measuring the
evaporative loss of trapped atoms from a finite-depth optical trap.
By using an ultracold atomic gas, with temperature $T \ll \hbar
\kappa / k_B$, the coherent amplification or damping of collective
motion [see Eq.\ \ref{eq:adaga}] may be neglected, and the atom
heating rate is related directly to the spectral density of photon
fluctuations in the cavity.

For this heating measurement, we prepared an ultracold gas of
$^{87}$Rb atoms within a high-finesse Fabry-Perot optical resonator
\cite{gupt07nonlinear} (Fig.\ \ref{fig:scheme}). One TEM$_{00}$ mode
of the cavity was excited resonantly with light at wavevector $k_t =
2 \pi / (850\,\mbox{nm})$. This light, far detuned from atomic
resonances, formed a one-dimensional optical lattice of depth
$U/k_{B} = 6.6(7) \, \mu$K in which the atoms were trapped and
evaporatively cooled to a temperature of $T
 = 0.8 \, \mu$K, as determined by time-of-flight measurements after the atoms
were released from the trap.  The atoms occupied approximately $
300$ adjacent sites in the optical lattice.  Given $k_B T < \hbar
\omz$, where $\omz = 2 \pi \times 42$ kHz is the axial trap
frequency in each lattice site, all axial vibration modes, including
the collective mode pertinent to cavity-based position measurements,
were cooled to their ground state. The atomic sample was probed
using light with wavevector $k_p = 2 \pi / (780\,\mbox{nm})$ that
was nearly resonant with another TEM$_{00}$ mode of the cavity.  For
this light, the cavity mirrors, separated by 194 $\mu$m and each
with 5 cm radius of curvature, displayed measured losses and
transmissions per reflection of 3.8 and 1.5 ppm, respectively,
yielding $\kappa = 2 \pi \times 0.66$ MHz.  The bare-cavity
resonance frequency for this mode $\omc$ was maintained at a
detuning of $|\Delta_{ca}|=
 2\pi\!\times\! (30 - 100)$ GHz from the $^{87}$Rb D2 atomic resonance.
The cavity was stabilized by passive \emph{in vaccuo} vibration
isolation and by active feedback based on transmission measurements
of the trapping light at wavevector $k_t$.

The atom-cavity coupling frequency $g_0\!=\!2\pi\!\times\! 14.4$ MHz
was determined from measured cavity parameters and by summing over
all excitations from the $|F = 1, m_F = -1\rangle$ hyperfine ground
state by $\sigma^+$ probe light on the D2 resonance line.  With the
atomic resonance half-linewidth being $\Gamma = 2 \pi \times 3$ MHz,
the single-atom cooperativity of $C = g_0^2 / 2 \kappa \Gamma = 52$
satisfies the criterion for strong coupling.

To measure the backaction heating near the cavity resonance, $N =
10^5$ atoms were loaded into the cavity, causing the cavity
resonance to be shifted by $\Delta_N\! =\! 2 \pi\! \times\! 100$ MHz
at the atom-cavity detuning of $\Delta_{ca}\! =\! 2 \pi\! \times\!
100$ GHz.  The cavity was then driven with probe light detuned by
$\dpc = 2 \pi \times 40$ MHz from the bare-cavity resonance.
Transmission through the cavity was monitored using single-photon
counting devices. The cavity photon number $\nbar$ was obtained from
the transmission signal using the measured quantum efficiency of
$0.040(8)$ for detecting intracavity photons.

\begin{figure}
\includegraphics[angle = 0, width = .48\textwidth]{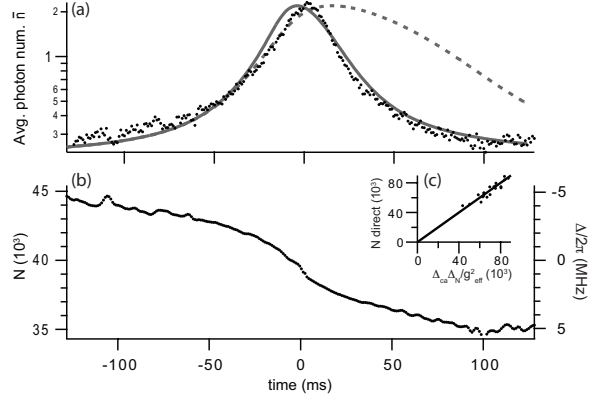}
\caption{Cavity-based observation of evaporative atomic losses due
to cavity-light-induced diffusive heating. (a) The intracavity
photon number, $\nbar$ (points, average of 30 measurements) is
monitored as the atom number is reduced by evaporation, and the
cavity resonance is brought across the fixed probe frequency.  The
expected $\nbar(t)$ excluding (dashed) or including (solid)
cavity-enhanced diffusive heating  are shown. (b) The atom number
$N(t)$ is inferred from  the measured photon number based on the
cavity lineshape. Atoms are lost at a background rate of 0.9(1)
s$^{-1}$ per atom away from the cavity resonance, and thrice faster
near resonance.  (c) The relation between
$2\Delta_{ca}\Delta_N/g_{0}^{2}|_{Z=0}$ and the atom number measured
directly by absorption imaging matches with predictions (line).}
\label{fig2}
\end{figure}

While the transmitted probe intensity was initially negligible owing
to the large detuning between the probe and cavity resonance
frequencies, the ongoing loss of atoms from the optical trap
eventually brought the atoms-cavity resonance near the probe
frequency, leading to discernible transmission (Fig.\
\ref{fig2}(a)).  We used this transmission signal to determine the
atom number $N$ and its rate of change $dN/dt$ as functions of time.
We related $\Delta_N$ to the instantaneous transmitted probe power
by assuming a Voigt lineshape for the cavity transmission with a
Gaussian kernel of rms frequency width
$\sigma\!=\!2\pi\!\times\!1.1$ MHz chosen to account for broadening
 due to technical fluctuations in
$\dpc$.  We also modified the lineshape to account for the
probe-induced displacements of the collective position $Z$ that were
as high as 3.5 nm for the maximum cavity photon number ($\nbar =
1.9$) used here \cite{gupt07nonlinear}. As shown in Fig.\
\ref{fig2}(b), the atom loss rate was strongly enhanced near
resonance due to increased light-induced heating.

\begin{figure}[tb]
\includegraphics[angle = 0, height = 1.35 in]{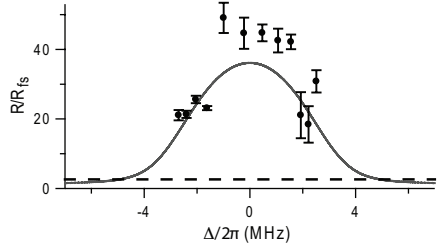}
\caption{Cavity-heating of a collective atomic mode in a strongly
coupled Fabry-Perot cavity over spontaneous emission dominated free
space heating.  The measured ratio $R/\rfs$ is shown with 1$\sigma$
statistical error bars.  For each measurement, a 12 ms-long range of
the probe transmission measurement data was used to determine
$dN/dt$ and $\nbar$. Systematic errors, at a level of 23\% at the
cavity resonance, arise from uncertainty in the background loss
rate, the background light level, and overall photo-detection
efficiency. Grey line shows theoretical prediction (with no
adjustable parameters) as described in the text. Dashed line shows
an upper bound on the off-resonance heating rate based on
measurements at $\Delta_{ca}\!=\!2\pi\!\times\!29.6$ GHz and
$\Delta\!=\!2\pi\!\times\!40$ MHz.} \label{doverdfsplot}
\end{figure}

From the observed loss rate we determined the per-atom heating rate
of the trapped atomic sample as $R=-U d(\ln N)/dt$ (Fig.\
\ref{fig2}, \ref{doverdfsplot}). Atoms experiencing intracavity
intensity fluctuations of cavity-resonant light were heated at a per
atom rate that is $R / \rfs \simeq 40$ times larger than that of
atoms exposed to a standing wave of light of \emph{equal intensity}
in free space. The cavity-induced heating was abated for light
detuned from the cavity resonance. While this cavity-enhanced
diffusion has been inferred from the lifetime \cite{maun05} and
spectrum \cite{munst99dyn} of single atoms in optical cavities, our
measurements are performed under experimental conditions that allow
its direct quantification.

To compare the observed heating rate with that expected based on
quantum-measurement backaction (Fig.\ \ref{doverdfsplot}), we
account for technical fluctuations in the probe detuning $\delta$ by
a convolution of the predicted frequency dependence with the
aforementioned Gaussian kernel.  For $\delta = 0$, this convolution
reduces the measured heating rate per photon by a factor of 0.7
below what is expected in absence of technical fluctuations. The
measured atom heating rates agree well with their predicted value,
confirming that the backaction heating of the atomic ensemble is at
the level required for quantum-limited measurements. Using the
relation between $\rcav$ and $\snn$, the measured heating rate may
be interpreted as a measurement of the spectral density of
intracavity photon number fluctuations where the atomic ensemble is
used as a mechanical sensing medium for these fluctuations.  From
the measured maximum heating rate of $R/\rfs = 43(10)$, the error
being predominantly systematic, and accounting for the convolved
cavity line shape, we obtain the spectral noise power of photon
fluctuations in a resonantly driven cavity as $S_{nn}/\nbar = 4.0(9)
\times 10^{-7}$ s, in agreement with the predicted $S_{nn}/\nbar =
2/ \kappa = 4.8 \times 10^{-7}$ s.

We have shown that heating due to cavity-induced fluctuations of the
optical dipole force dominates the heating of a trapped atomic gas
near resonance. To highlight this finding further, we measured the
atom heating rate due to intracavity light that is far from the
cavity resonance, for which one should observe the
spontaneous-emission-dominated heating of atoms in free space. For a
atom-cavity detuning of $\Delta_{ca}\! =\! 2\pi\! \times\! 29.6$ GHz
and $N \simeq 9000$ atoms, we excited the cavity for a variable time
with probe light at detuning $\Delta = 2 \pi \times 40$ MHz with an
intracavity photon number of $\nbar = 2$.  From the decay rate of
$N$, we observed a probe-light-induced per-atom loss rate that, if
ascribed completely to diffusive heating of the atomic sample,
yields a heating rate of $R/\rfs\!=\! 2.9(7)$, far smaller than that
observed at the cavity resonance. Yet, these losses exceeded those
expected based on diffusion from Rayleigh scattering. This
discrepancy may be explained by additional effects of Raman
scattering.  Atoms scattered by the $\sigma^+$ probe light into
different hyperfine ground states couple to the cavity probe light
with different strength, thereby changing the relationship between
$\Delta_N$ and the atom number $N$.  These additional effects appear
sufficient to account for our observations, yet  are constrained by
our measurements to contribute only slightly to the atom losses
observed from probe light at the cavity resonance.

This work demonstrates the bright prospects for studying quantum
aspects of the motion of macroscopic ($\neff \simeq 10^5$ atoms)
mechanical systems. The optical confinement of ultracold atoms
within a high-finesse optical resonator enabled the construction of
a nearly ground state mechanical resonator.  The identification, and
quantitative measurement of cavity heating as the quantum backaction
of position measurement on a collective mode of a mechanical object
is a significant demonstration of the ultracold-atom approach to
quantum micro-mechanics.

Working in the non-granular regime and with an atomic medium at
sufficiently low temperatures so that cavity cooling/anti-cooling
could be neglected, we may interpret the measured backaction heating
as a direct measurement of the spectrum of photon fluctuations in a
driven cavity, a quantity of fundamental interest in quantum optics.
We note that these fluctuations are not visible in the coherent
light transmitted through the cavity, for which the shot-noise
spectrum remains white (see supplemental information). Specifically,
in a cavity driven by coherent laser light, the atoms serve as an
\emph{in-situ} heterodyne detector of cavity-enhanced fluctuations
of the electromagnetic field, with the two quadratures of collective
motion serving as two heterodyne receivers at the beat frequency
$\omega_{z}$. At present, by quantifying only the total heating rate
of the trapped atomic gas, we cannot access information on the
individual noise quadratures. However, augmented by time-resolved
measurements of the collective motion, as demonstrated in Ref.\
\cite{gupt07nonlinear}, our setup may also serve to probe
quadrature-squeezed light before the intracavity squeezing is
degraded by attenuation outside the cavity \cite{loud87}.

We have demonstrated that atoms in a strong-coupling cavity are
heated optically at a rate that exceeds that calculated for
free-space illumination.  This fact presents a challenge to
cavity-aided non-destructive measurements of atom number or spin
with uncertainty below the standard quantum limit
\cite{kuzm98,taka99squeezing,bouc02squeezing,auzi04,hald99,geremia04}.
In such measurements, the sensitivity gained by increasing the probe
light fluence is eventually offset by the increased disturbance of
the atoms due to incoherent light scattering.  Our work suggests
that cavities with single-atom cooperativity beyond $C = 1$ will
yield benefits to these measurements only if the measurement is made
insensitive to the atomic position, e.g.\ by placing atoms at
antinodes of the cavity field or in traps for which $\omega_{z} \gg
\kappa$.

We thank T.\ Purdy and S.\ Schmid for early contributions to the
experimental apparatus, and S.M.\ Girvin, J.\ Harris, H.J.\ Kimble,
H.\ Mabuchi, and M.\ Raymer for helpful discussions.  This work was
supported by AFOSR, DARPA, and the David and Lucile Packard
Foundation.

\clearpage

\begin{center}\textbf{\Large{Supplementary Information}}\end{center}

\begin{center}\textbf{Development of collective modes}\end{center}

The atoms in our system are confined at many locations within a one
dimensional optical lattice of wavevector $k_{t} = 2 \pi /850$ nm,
yet interact with the cavity mode at the position dependent coupling
rate $g(z) = g_{0} \sin k_{p}z$.  Because of the position dependent
coupling and the distribution of the atoms, a single collective
degree of freedom (but not a center of mass degree of freedom)
interacts with the cavity mode.

We consider  that each atom is trapped harmonically with frequency
$\omz$ and trap center $\zbari$, where we denote the displacement of
atom $i$ from its trap's center by the operator $\delta z_i = z_i -
\zbari = \zho (\ahat_i^\dagger + \ahat_i)$ with $\zho = \sqrt{\hbar
/ 2 m \omz}$ being the harmonic oscillator length and atom field
operators $\ahat_i$ and $\ahat_i^\dagger$ conventionally defined. We
assume that the atomic displacements are small ($k_p \delta z_i \ll
1$) and that the cavity-atom detuning is large ($|\dca| \gg g_0
\sqrt{N}$). Omitting some constant terms, we obtain a Hamiltonian
describing the coupled atoms/cavity system as
\begin{equation}
\hami = \left(\hbar \omc + \sum_i \left[ \frac{\hbar
g^2(\zbari)}{\dca} - f_i \delta z_i\right] \right) n + \hami_a+
\hami_{in} \label{eq:hami},
\end{equation}
where $n$ is the cavity photon number operator, $\hami_a = \sum_i
\hbar \omz \ahat_i^\dagger \ahat_i$, and $\hami_{in}$ describes
optical modes external to the cavity and their coupling to the
cavity field \cite{wall94book}. Here, the per-atom cavity resonance
shift is expanded to first order in the atomic position operator,
with $f_i = -\hbar \partial_z g^2(z) / \Delta_{ca} = f_0 \sin(2 k_p
\zbari)$ being the optical dipole force on atom $i$ from a single
cavity photon.

We define a collective position operator $Z = (\neff)^{-1} \sum_i
\sin(2 k_p \zbari) \delta z_i$, and the conjugate momentum $P =
\sum_i \sin(2 k_p \zbari) p_i$, with $p_i$ being the momentum of
atom $i$, and $\neff = \sum_i \sin^2(2 k_p \zbari)$.  The cavity
then serves to monitor a specific collective mode of motion in the
atomic ensemble, with the cavity resonance being shifted by
$\Delta_N - \neff\  f_0 Z/ \hbar$ where $\Delta_N = \sum_i
g(\zbari)^2/\dca$ is the cavity frequency shift with all atoms
localized at their potential minima.

Given these collective operators we can write equations of motion;
\begin{align}
\dot{P} &= \sum_{i} \sin (2k_{p}\zbari) \Bigl(-m \omz^{2} \delta z_i + f_{i} n\Bigr)\\
&=-\neff\ m \omz^{2} \Bigl(Z - \frac{f_{0} \nbar}{m \omz^{2}} \Bigr) + f_{0}(n-\nbar).\\
\dot{Z} &= \frac{P}{\neff\ m}.
\end{align}

A constant average optical force of $\nbar$ cavity photons displaces
the collective position variable by $\Delta Z = (\hbar k g_0^2 / m
\omz^2 \dca) \nbar$ and thereby shifts the cavity resonance
frequency to $\omcp = \omc + \Delta_N - \neff f_0 \Delta Z / \hbar$.
We define collective quantum operators $a$ and $a^\dagger$ through
the relations $Z - \Delta Z = \Zho (a^\dagger + a)$ and $P = i \Pho
(a^\dagger - a)$, with $\Zho = \zho/\sqrt{\neff}$ and $\Pho = \hbar
/ (2 \Zho)$.  With these substitutions, we have the Hamiltonian
describing the collective mode--cavity system:
\begin{eqnarray}
\hami = \hbar \omcp n - \neff f_{0} \Zho (\adag +a) (n-\nbar) +
\hbar \omz \adag a + \hami_{in} \label{eq:colhami}.
\end{eqnarray}

\begin{center}\textbf{Calculation of the heating rate}\end{center}

Given equation \ref{eq:colhami}, we can draw directly on existing
results which analyze cavity cooling and heating for similar
Hamiltonians \cite{marq07sideband}.  For clarity, however, we
present a derivation of the heating rate below. From Eq.\
\ref{eq:colhami} we obtain equations of motion for $a$ and for the
cavity field operator $b$.
\begin{eqnarray}
\frac{d a}{d t} & = & - i \omz a + i \kappa \epsilon (n-\nbar) \label{eq:dadt2},\\
\frac{d b}{d t} & = & -i \omcp b + i \kappa \epsilon (\adag + a) b -
\kappa b + \sqrt{2 \kappa} b_{in} \label{eq:dbdt2},
\end{eqnarray}
where $\kappa$ is the decay rate of the cavity field and $b_{in}$
represents the coherent-state input field that drives the cavity.
We have introduced the granularity parameter $\epsilon = \neff f_{0}
\Zho/ (\hbar \kappa)$ as discussed in the text.  We can now express
the atomic field operator as,
\begin{align}
a(t) = e^{-i\omz t} a(0) + i \kappa \epsilon \int_{0}^{t}dt' e^{-i
\omz (t-t')}\bigl(n(t') -\nbar \bigr) \label{eq:atomfield}.
\end{align}
From here, we evaluate the rate of change of the atomic energy:
\begin{align}
\frac{d}{dt} (\adag a) &= \Bigl(\frac{d}{dt} \adag\Bigr)_{t}a(t) + \adag(t)\Bigl(\frac{d}{dt} a \Bigr)_{t}\\
&=\Bigl[ i \omz \adag (t) - i \kappa \epsilon (n(t)-\nbar)\Bigr] a(t) + \adag(t) \Bigl[-\omz \adag(t) + i \kappa \epsilon (n(t)-\nbar) \Bigr] \\
&=2 \kappa^{2} \epsilon^{2}\  \textrm{Re}\Biggl[ \int_{0}^{t}dt' (n(t)-\nbar) (n(t')-\nbar) e ^{-i \omz (t-t')}\Biggr] \nonumber \\
&\quad \quad \quad \quad \quad+ i \kappa \epsilon \Bigl(\adag(0)
(n(t)-\nbar) e^{i\omz t} -(n(t)-\nbar) a(0) e^{-i \omz t} \Bigr)
\label{eq:heat1}.
\end{align}

For the sake of evaluating the cavity field evolution we restrict
our treatment to times which are short compared to the timescale
over which the atomic motion is significantly varied by interaction
with the light. Under this ansatz we approximate Eq.\
\ref{eq:atomfield} as
\begin{align}
a(t) \simeq e^{-i \omz t} a(0) \label{eq:aapprox}.
\end{align}
Inserting this solution for the atomic field operator into the
equation of motion for the cavity field, (\ref{eq:dbdt}) we have the
following for the frequency components of $b$:
\begin{align}
-i \omega b(\omega) = -i \omcp b(\omega) - \kappa b(\omega) +
\sqrt{2\kappa} b_{in} (\omega) + i \kappa \epsilon \bigl(a(0)
b(\omega-\omz) + \adag (0) b  (\omega+\omz)\bigr).
\end{align}
Defining $L(\omega) = (1-i(\omega - \omcp)/\kappa)^{-1}$, we obtain
\begin{align}
b(\omega) = \frac{L(\omega)}{\kappa} \Bigl[ \sqrt{2 \kappa}
b_{in}(\omega) + i \epsilon \Bigl(a(0) b(\omega - \omz) +
\adag(0)b(\omega+\omz)\Bigr)\Bigr].
\end{align}
We can solve this equation iteratively,
\begin{align}
b(\omega) = \frac{L(\omega)}{\kappa} \Bigl[ \sqrt{2 \kappa}
b_{in}(\omega) + i \epsilon \sqrt{2\kappa}\Bigl(a(0) L(\omega -
\omz) b_{in}(\omega - \omz)\nonumber\\ + \adag(0)L(\omega +
\omz)b_{in}(\omega+\omz)\Bigr) + \mathcal{O}\bigl(|\epsilon
a(0)|^{2}\bigr)\Bigr].
\end{align}
In the non-granular regime $\epsilon \ll 1$, and assuming small
values of $a(0)$, i.e.\ that the atoms are sufficiently cold, we
neglect terms of order $\epsilon^3$ or higher.

Returning to Eq.\ \ref{eq:heat1} we now have
\begin{align}
n(t)&= \frac{1}{2\pi} \int d\omega_{1}\ d\omega_{2}e^{i(\omega_{1} - \omega_{2})t}\  \bdag(\omega_{1})b(\omega_{2})\\
&= \frac{1}{2\pi} \int d\omega_{1}\ d\omega_{2}e^{i(\omega_{1} -
\omega_{2})t}\frac{ L^{*}(\omone)
L(\omtwo)}{\kappa^{2}}2\kappa\Biggl[\bdag_{in}(\omone)b_{in}(\omtwo)
+ \nonumber
\\
& \quad i \epsilon \bdag_{in}(\omone)  \Bigl(a(0) L(\omtwo - \omz)
b_{in}(\omtwo - \omz) + \adag(0)L(\omtwo +
\omz)b_{in}(\omtwo+\omz)\Bigr)- \nonumber
\\
& \quad  i \epsilon  \Bigl(\adag(0) L^{*}(\omone - \omz)
b^{\dagger}_{in}(\omone - \omz) +a(0)L^{*}(\omone +
\omz)b^{\dagger}_{in}(\omone+\omz)\Bigr)b_{in}(\omtwo) \Biggr].
\end{align}
With the above normally ordered product of operators $b_{in}$ we are
justified in replacing:
\begin{align}
b_{in}(\omega) \rightarrow \sqrt{\pi \kappa \nmax} \, \delta(\omega-\omega_{p}),\\
\bdag_{in}(\omega) \rightarrow \sqrt{\pi \kappa \nmax}\,
\delta(\omega-\omega_{p}),
\end{align}
where $\omega_{p}$ is the frequency of a probe laser, and $\nmax$ is
the maximum intracavity photon number for resonant cavity
excitation.  Finally, we obtain
\begin{align}
n(t) =& \nbar\Bigl[1+ i\epsilon \Bigl( a(0) L(\omega_{p} + \omz) e^{-i\omz t} + \adag(0) L(\omega_{p}-\omz)e^{+i\omz t}\Bigr)- \nonumber\\
& \quad \quad \quad i \epsilon\Bigl(\adag(0)
L^{*}(\omega_{p}+\omz)e^{+i\omz t} +  a(0) L^{*}(\omega_{p} - \omz)
e^{-i\omz t}\Bigr)\Bigr].
\end{align}
Here we have substituted $\nbar = \nmax |L(\omega_{p})|^{2}$.

We are now in a position to evaluate the heating rate:
\begin{align}
\frac{d}{d t} E &= \hbar \omz \Bigl\langle \frac{d}{dt} \adag
a\Bigr\rangle  \\&= 2 \hbar \omz \kappa^{2} \epsilon^{2}
\textrm{Re}\Biggl[ \int_{0}^{t}dt' \langle
\bigl(n(t)-\nbar\bigr)\bigl( n(t')-\nbar \bigr)\rangle e ^{-i \omz
(t-t')}\Biggr] +\nonumber \\ &\quad \quad \quad \quad \quad \quad
\quad i  \hbar \omz  \kappa \epsilon \Bigl\langle\adag(0)
\bigl(n(t)-\nbar\bigr) e^{i\omz t} -\bigl(n(t)-\nbar\bigr) a(0)
e^{-i \omz t} \Bigr\rangle .\label{eq:totalheatnoeval}
\end{align}
Addressing the first term first;  for a linear cavity driven by a
constant coherent state input, we substitute the relation,
\begin{align}
\langle n(\tau) n(0) \rangle -\langle  n(\tau)\rangle^{2} = \nbar
e^{i (\omega_{p} - \omega_{c}') \tau - \kappa \tau}.
\label{eq:twotime}
\end{align}
Assuming the system is in a steady state, in that $\langle n(t)
n(t')\rangle = \langle n(t-t')n(0)\rangle$, and substituting
$\nbar^{2} = \langle n(\tau) \rangle ^{2}$ we obtain for the first
half of the heating rate,
\begin{align}
2 \hbar \omz \kappa \epsilon^{2} \nbar \Bigl(\frac{1}{1+ (\omega_{p}
- \omega_{c}' - \omz)^{2}/\kappa^{2}}\Bigr) =  \hbar \omz \kappa^{2}
\epsilon^{2}[ \snnminus(\omz)].
\end{align}
Here we have introduced the spectral density of photon number
fluctuations $\snn^{(\pm)}(\omega)= 2 \nbar \kappa  (\kappa^{2} +
(\Delta \pm \omega)^{2})^{-1}$ \cite{marq07sideband}, with $\Delta =
\omega_{p} - \omc'$ begin the probe detuning from the atoms shifted
cavity resonance.

The second term in Eq.\ \ref{eq:totalheatnoeval} accounts for the
effect of transient atomic motion on the cavity field.   To evaluate
this term we take the time average over an atomic oscillation
period.
\begin{align}
 &i \kappa \epsilon \Bigl\langle\adag(0) \bigl(n(t)-\nbar\bigr) e^{i\omz t} -\bigl(n(t)-\nbar\bigr) a(0) e^{-i \omz t} \Bigr\rangle\\
 &=\nbar \epsilon^{2} \kappa \Bigl(L(\omega_{p} + \omz) - L^{*}(\omega_{p} - \omz) +L(\omega_{p} - \omz) - L^{*}(\omega_{p} + \omz) \Bigr)\langle \adag(0)a(0)\rangle\\
 &=\kappa^{2}  \epsilon^{2} \bigl[\snnminus(\omz) - \snnplus(\omz)\bigr] \langle \adag a\rangle.
\end{align}
These terms represent cavity cooling/anti-cooling.  In total, the
change in energy is,\begin{align} \frac{d}{dt} E =   \hbar \omz
\kappa^{2} \epsilon^{2} \Biggl[ \snnminus(\omz) +
\Bigl(\snnminus(\omz) - \snnplus(\omz)\Bigr)\langle \adag a\rangle
\Biggr].
\end{align}

\begin{center}\textbf{Measuring backaction heating by the evaporative loss of trapped atoms}\end{center}

The accuracy of our measurement depends on assumptions made in
interpreting the observed transmission lineshapes, several of which
we verified experimentally.  For example, we examined the dynamics
of evaporative cooling in the atomic medium.  For this, we
interrupted the cavity transmission measurement, released the atoms
from the intracavity  optical trap and imaged them 4 ms later to
measure their temperature. Within our measurement resolution of 0.1
$\mu$K, this temperature remained constant. Thus, our quantification
of heating through the rate of atom loss is valid. Furthermore, by
extinguishing the cavity probe light momentarily during cavity
probing, and comparing the cavity transmission when the probe was
turned off and then turned on again, we determined a timescale of
$3$ ms for $N$ to equilibrate by evaporative cooling following an
increase of thermal energy of the collective mode.  Since this
timescale is short compared to the $\simeq 100$ ms span of the
resonant transmission signal, we are justified in using simultaneous
measurements of $dN/dt$ and $\nbar$ to determine the instantaneous
heating rate.

To interpret our measurements as relating to the quantum nature of
the intracavity field, it was necessary to establish that quantum
fluctuations dominate over classical, technical intensity
fluctuations which would also lead to heating \cite{gehm98heat}. For
this, we measured the light-induced heating for varying probe
intensities, with $\nbar$ at the cavity resonance ranging from
$\nbar = 0.2$ to 20.  Noting that the contribution of quantum
fluctuations to the atom heating rate scales as $\nbar$ while that
of technical fluctuations scales as $\nbar^2$, we find that
technical fluctuations account for less than 10\% of the atom
heating rate at the light level used for Figs.\ 2 and 3.

\begin{center}\textbf{Visibility of photon fluctuations outside the cavity}\end{center}

In this section, we provide support for the
well-established\cite{car91open} but oft-forgotten result that the
spectrum of intra-cavity quantum fluctuations of the photon number
is not visible in light transmitted through the coherently driven
cavity. For a two sided cavity we have \cite{wall94book}:
\begin{align}
\cout(t) + \cin(t) &= \sqrt{\kappa} b(t),\\
\dout(t) + \din(t) &=  \sqrt{\kappa} b(t),\\
b(\omega) &=  \frac{\sqrt{\kappa} \cin(\omega) +  \sqrt{\kappa} \din
(\omega)}{\kappa - i (\omega - \omega_{c}')}.
\end{align}
The operators $\cout, \dout, \cin, \din$ are photon annihilation
operators for the outgoing and in-going fields on either side of the
cavity, and $b$ is again the cavity field annihilation operator. The
known commutation relations are,
\begin{align}
\Bigl[\cin(\omega_{1}), \cindag(\omega_{2})\Bigr] =
\delta(\omega_{1} - \omega_{2}),\quad \quad \Bigl[\cin(t_{1}),
\cindag(t_{2})\Bigr] = \delta(t_{1} -t_{2}),
\end{align}
and similarly for $\din$.  To examine the spectrum of photon
fluctuations inside the cavity  we calculate the commutation
relation for the cavity field operator:
\begin{align}
\Bigl[b(t_{1}), \bdag(t_{2})\Bigr] &= \int \frac{d\omega_{1}\ d
\omega_{2}}{2 \pi} \frac{e^{-i\omega_{1} t_{1}}}{\kappa-i
(\omega_{1} - \omega_{c}')} \frac{e^{i\omega_{2} t_{2}}}{\kappa+i
(\omega_{2} - \omega_{c}')}
\times \nonumber\\
&\quad \quad \quad  \quad\Bigl[\sqrt{\kappa} \cin(\omega_{1}) +  \sqrt{\kappa} \din (\omega_{1}),\sqrt{\kappa} \cindag(\omega_{2}) +  \sqrt{\kappa} \dindag (\omega_{2})\Bigr]\\
 &= \int \frac{d\omega_{1}\ d \omega_{2}}{2 \pi}
\frac{e^{-i\omega_{1} t_{1}}}{\kappa-i (\omega_{1} - \omega_{c}')}
\frac{e^{i\omega_{2} t_{2}}}{\kappa+i (\omega_{2} - \omega_{c}')}
2\kappa \delta(\omega_{1}- \omega_{2})
\\
&= \int \frac{d \omega}{2 \pi} \frac{2\kappa}{\kappa^{2} + (\omega-\omega_{c}')^{2}} e^{i\omega (t_{2}-t_{1})}\\
&=e^{i \omega_{c}' (t_{2}-t_{1}) - \kappa|t_{2}-t_{1}|}
.\label{eq:cavcom}
\end{align}
From this we obtain the two-time correlation in Eq.\
\ref{eq:twotime}.  Now, for the cavity output, (say, $\dout$),
\begin{align}
\Bigl[\dout(t_{1}), \doutdag(t_{2})\Bigr] &= \int \frac{d\omega_{1}\ d\omega_{2}}{2\pi} e^{i\omega_{1}t_{1} + i \omega_{2}t_{2}}\Biggl[ \frac{\kappa \cin(\omega_{1}) +  \kappa \din (\omega_{1})}{\kappa - i (\omega_{1} - \omega_{c}')} - \din(\omega_{1}),\nonumber\\ &\quad \quad \quad \quad \quad \quad \quad \quad \quad \quad \quad \quad \quad \quad \quad  \frac{\kappa \cindag(\omega_{2}) +  \kappa \dindag (\omega_{2})}{\kappa + i (\omega_{2} - \omega_{c}')} - \dindag(\omega_{2}) \Biggr] \\
&= \int \frac{d\omega_{1}\ d\omega_{2}}{2\pi} e^{i\omega_{1}t_{1} +
i \omega_{2}t_{2}} \Biggl(
 \frac{2 \kappa^{2} \delta(\omega_{1}-\omega_{2})}{(\kappa - i (\omega_{1} - \omega_{c}'))(\kappa + i (\omega_{2} - \omega_{c}'))} -\nonumber \\ &\quad \quad \quad \quad \quad \quad \quad \quad \quad  \quad \quad \quad
  \frac{\kappa \delta(\omega_{1} - \omega_{2})}{\kappa - i (\omega_{1} - \omega_{c}')}- \frac{\kappa \delta(\omega_{1} - \omega_{2})}{\kappa + i (\omega_{2} - \omega_{c}')}\Biggr)\\
  &=
 \int \frac{d \omega}{2 \pi } e^{i \omega (t_{2}-t_{1})} = \delta(t_{1} - t_{2}).
\end{align}
The commutation relations for fields outside the cavity are the same
as for light entering the cavity, and do not carry any evidence of
the photon number dynamics (Eq.\ \ref{eq:cavcom}) inside the cavity.

\bibliographystyle{apsrev}

\end{document}